\documentclass[useAMS,usenatbib]{mn2e}

\usepackage{graphicx}
\usepackage{amssymb,amsmath}

\title[Stability of the triangular Lagrangian points]{Stability chart of the triangular points in the elliptic restricted problem of three bodies}
\author[T. Kov\'acs]{T. Kov\'acs$$\thanks{E-mail: tkovacs@konkoly.hu}\\
  Konkoly Observatory, Research Centre for Astronomy and Earth Sciences, Hungarian Academy of Sciences,\\P.O. Box 67, H-1525  Budapest, Hungary}

\begin{document}

\date{Accepted -- Received --}

\pagerange{\pageref{firstpage}--\pageref{lastpage}} \pubyear{2002}

\maketitle

\label{firstpage}

\begin{abstract}
  The possible observations of Trojan-like extrasolar planets stimulate the deeper understanding of the stability behaviour of the co-orbital resonant motion. By using Hill's equations and the energy-rate method an analysis of the stability map of the elliptic restricted three-body problem is performed. Regions of the $\mu-e$ parameter plane are described numerically and related to the resonant frequencies of librational motion. Stability and instability can therefore be obtained by analysing the two independent frequency modes depending on system parameters. The key role of the long period libration in determining the structure of the stability is demonstrated and also a stability mechanism is found that is responsible for extended life time of the test particle in the unstable domain of the stability map.
\end{abstract}


\section{Introduction}
The stability of the Lagrangian triangular points $L_4$ and $L_5$ in the elliptic restricted three-body problem (ERTBP) is an extensively studied topic of celestial mechanics. Recent directions in extrasolar planetary dynamics require to rethink our knowledge about the historical results of pure dynamical astronomy. The stability of motion around $L_4$ and $L_5$ depends on two parameters in the ERTBP, namely on the orbital eccentricity ($e$) and the mass parameter ($\mu$) of the primaries. 

The main results on the stability of $L_4$ and $L_5$  were published in \citet{Dan}, \citet{Ben}, and \citet{Tsch}. Numerical studies about the stability was carried out by \citet{Loh} and \citet{Mar}. It was shown by \citet{Erd1,Erd2} and \citet{Raj} that there are secondary resonances responsible for the fine structure of the stability chart. Recently \citet{Sch} demonstrated the stability phenomena in the spatial problem, too.

In this work we present a fast and accurate method to construct the detailed stability map of the problem using Hill's equations. The results reveal the complex structure of the $\mu-e$ plane and give evidence for peaks of longer escape times in the unstable region of the parameter space.

\section[]{Equations of motion and methods}
The behaviour of the test particle near the $L_4$ and $L_5$ points in the planar elliptic restricted three-body problem can be described by the well-known equations of variation \citep{Szeb}
\begin{subequations}
  \label{ertbp}
  \begin{align}
    x''-2y'&=rc_{1}x,\label{ertbpa}\\
    y''+2x'&=rc_{2}y\label{ertbpb}
  \end{align}
\end{subequations}

where
\begin{equation*}
  \label{rtbp_aux}
  \begin{split}
    r&=(1+e\cos v)^{-1},\\
    \mu&=\frac{m_{2}}{m_{1}+m_{2}},\\
    g&=3\mu(1-\mu),\\
    c_{i}&=1.5(1+(-1)^{i}\sqrt{1-g}).\hspace{10mm}(i=1,2)
  \end{split}
\end{equation*}
In Eqs. \eqref{ertbp} the prime denotes the derivative with respect to the true anomaly $v.$  The orbital parameter $e$ and the dimensionless mass parameter $\mu$ are involved in the problem, $m_1$ and $m_2$ are the masses of the primaries.
\subsection{Hill's equations}
By a suitable transformation \citep{Tsch}, the 4-th order system \eqref{ertbp} can be separated into two independent 2-nd order systems. From these we can derive an equivalent set of Hill's equations \citep{Mei1,Mei2}:
\begin{subequations}
  \label{hill}
  \begin{align}
    \xi_{1}''+J_{1}(v;e,\mu)\xi_{1}&=0,\label{hilla}\\
    \xi_{2}''+J_{2}(v;e,\mu)\xi_{2}&=0,\label{hillb}
  \end{align}
\end{subequations}
where $J_{i}(v;\mu,e)$ are $2\pi$-periodic functions containing the system parameters. Therefore the system \eqref{hill} can be thought as of two decoupled harmonic oscillators with variable frequencies. The characteristic units in Eqs.~\eqref{ertbp} and \eqref{hill} were chosen so that the primaries' period is $2\pi,$ i.e. the mean motion is 1. For a detailed description of Eq.~\eqref{hill} see Appendix A.
\subsection{Energy-rate method}

The energy-rate method (ERM) \citep{Jaz} gives the opportunity, beside the well-known methods (Floquet theory, Poincar\'e-Linstedt series), to investigate the stability boundary of a system
\begin{equation}
  \label{spring}
  \ddot x + f(x)+g(x,\dot x,t)=0,
\end{equation}
where
\begin{equation}
  \label{spring_cond}
  g(0,0,t)=0
\end{equation}
and $f(x)$ is a single variable function, and $g(x,\dot x,t)$ is a nonlinear periodic function of the time $t.$ The functions $f$ and $g$ may depend on some parameters of the system. It is trivial from Appendix A that Eqs.~\eqref{hill} can be written in the form of Eq.~\eqref{spring}
\begin{equation}
  \label{hill_spring}
  \xi_{i}''-2\xi_{i}-\overline{J_{i}}(v;\mu,e)\xi_{i}=0,\hspace{10mm}(i=1,2)
\end{equation}
where
\begin{equation}
  \label{Jbar}
  \overline{J_{i}}(v;\mu,e)=rc_{1}-\frac{3Q_{i}+c_{2}}{q_{12}^{(i)}}+3\left[\frac{q_{22}^{(i)}}{q_{12}^{(i)}}\right]^{2}.
\end{equation}
$\overline{J_i}$ corresponds to excitation $g(x,\dot x,t)$ in Eq.~\eqref{spring} as one may check in Eqs.~\eqref{apphill}.

As is well-known if any non-conservative force is present in the system, the mechanical energy does not remain constant. In other words, due to the effect of $g(x,\dot x, t)$ in Eq.~\eqref{spring} for a given set of parameters, energy is either pumping into or withdrawing from the system. It can be verified \citep{Jaz} that in a system such as \eqref{spring} the energy integral has the following form
\begin{equation}
  \label{dotE}
  \dot E=\frac{\mathrm{d}}{\mathrm{d}t}(E)=\frac{\mathrm{d}}{\mathrm{d}t}\left(\frac{1}{2}\dot{x}^2+\int f(x)\dot x\,\mathrm{d}t\right)=-\dot xg(x,\dot x,t)
\end{equation}
where $T(\dot x)=\dot{x}^2/2$ is the kinetic and $V(x)=\int f(x)\,\mathrm{d}x$ is the potential energy for free motion.

It is also known that the time derivative of the mechanical energy must be zero over one period for conservative and autonomous systems in a steady state periodic cycle. In order to find the parameters that belong to stable or unstable solutions we may evaluate the energy over a period 
\begin{equation}
  \label{Eav}
  E_{\rm{av}}=\frac{1}{T}\int_0^T\dot{E}\,\mathrm{d}t
\end{equation}
where $T$ is the period of the external force $g.$ If $E_{\rm{av}}>0,$ provided that $f$ is a restoring force, then energy is inserted into the system, and Eq.~\eqref{spring}) is unstable. But, if $E_{\rm{av}}<0,$ energy is subtracted and the system is stable. On the common boundary $E_{\rm{av}}=0.$
\section{Stability analysis}

In the following the stability chart of the $L_4$ and $L_5$ points is studied using the ERM. A common way in stability analysis is to identify the characteristic roots of the problem that characterize the dynamically different domains of the parameter plane. The ERM allows us, if the equations are in a suitable format, a simple and fast method to distinguish the sets of parameters corresponding to stable or unstable dynamics.

In order to get the classical stability map of the $\mu-e$ plane one has to compute Eq.~\eqref{Eav} numerically for both Eqs.~\eqref{hill_spring}
\begin{equation}
  \label{Eav_hill}
  E_{\rm{av}}^{(i)}=\frac{1}{2\pi}\int_0^{2\pi}\overline{J_{i}}(v;\mu,e)\xi'_{i}\,\mathrm{d}v\hspace{7mm}(i=1,2).
\end{equation}

Note that equations \eqref{hill} have a \textit{repulsive} force, $f(\xi)=-2\xi,$ (see \eqref{hill_spring}) responsible for free motion. Therefore the stable solution for some $(\mu,e)$ pairs can be realized when the averaged energy $E_{\rm{av}}>0.$ Consequently, unstable motion appears in the opposite case, $E_{\rm{av}}<0.$

Since Eqs.~\eqref{hill} are linear ODEs with periodic coefficients, the choice of initial conditions is arbitrary, i.e. they do not affect the stability of the solution. Throughout this work initial conditions $(\xi_i,\xi'_i)=(1,0)$ are used. The control parameter that measures the accuracy $E_{\rm{av}}=0,$ is set to $\epsilon=5\cdot 10^{-6}.$

\begin{figure}
  \centering
  \includegraphics[width=8cm]{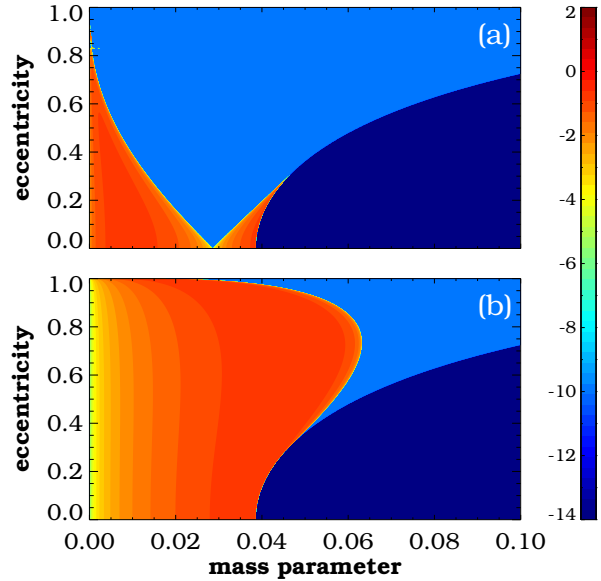}
  \caption{Stability map in the $\mu-e$ plane. Top: different features of motion around $L_{4}$ and ${L_5}$ governed by Eq.~\eqref{hilla}. Bottom: stability domains corresponding to Eq.~\eqref{hillb}. The contour bar indicates the magnitude of the average energy for given pairs of the parameters $(\mu,e).$  The integration time is $2\pi,$ one period of the primaries. The number of the parameter pairs is set to $10^6,$ i.e. 1000 in each direction with equal step size. (Colour figures are available online.)}
  \label{fig:en_2pi}
\end{figure}

Figure~\ref{fig:en_2pi}(a) depicts the classical stability map of the ERTBP obtained from Eq.~\eqref{hilla} by the ERM. The light gray region (orange online), where $E_{\rm{av}}>0$ for one period, corresponds to librational motion around $L_{4}$ and $L_{5}.$ The dark gray area (blue) represents the negative energy domain related to the set of parameters of unstable motion. The black part of the figure (dark blue), situated at the bottom right quarter, belongs to parameters where Eqs.~\eqref{hill} have no solutions. For better visualization (log-scale) we cut the positive energies at $10^2$ and set the negative values to $10^{-10},$ additionally, where the problem has no solution to $10^{-14}.$ Therefore, the contour bar is organized as follows: the light gray colour ($\mathrm{orange,}\;10^{-10}\leq E_{\rm{av}}\leq 10^{2}$) represents libration, the dark gray ($\mathrm{blue,}\;E_{\rm{av}}=10^{-10}$) denotes unstable motion, and the black domain ($\mathrm{dark\,blue,}\;E_{\rm{av}}=10^{-14}$) is for complex frequencies. The border of the stability domain is at $E_{\rm{av}}=0.$ The charecteristic roots of Eqs.~\eqref{ertbp} \citep{Ben} devide the stability chart exactly into the same parts.
  
  Figure~\ref{fig:en_2pi}(b) shows the stability chart provided by Eq.~\eqref{hillb}. It can be seen that the librational domain (light gray (orange online)) is more extended than in panel (a), however, the bottom right part (black/dark blue) is exactly the same. If we want to identify the stability region around the $L_4$ and $L_5$ points in the ERTBP determined by Hill's equations, we need the intersection of all the individual stable regions in Fig.~\ref{fig:en_2pi}(a) and \ref{fig:en_2pi}(b). One can immediately verify that the upper panel encompasses the domain in question. In other words, this shows that merely the first Hill's equation in Eqs.~\eqref{hill} is responsible for the well-known structure of the $\mu-e$ plane. The possible reason will be discussed in Section 5. Accordingly, we can state that the stability chart obtained from Eq.~\eqref{hill} is consistent with earlier analysis.
  
  One of many advantages of the ERM is that if we integrate the equation(s) of motion until $t=nT$ where $n$ is an integer number and $T$ is the period of the external force, new, so-called \textit{''splitting''} curves appear on the stability map. (The name comes from the fact that these resonant periodic lines split the stable regions into several parts.) Actually, these curves are \textit{super- and sub-harmonic oscillations} and describe higher order periodic responses of the parametric equations.
  
\begin{figure}
    \centering
    \includegraphics[width=8cm]{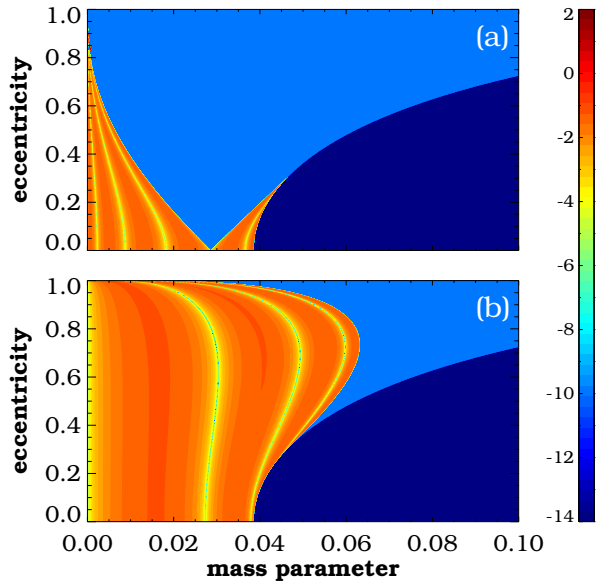}\\
    \caption{The $\mu-e$ plane for integration time $t=8\pi.$ Top and bottom: stability zones derived from Eqs.~\eqref{hilla} and \eqref{hillb}, respectively. Increasing the computation time by the multiplication of the excitation period, resonant curves turn up inside the stable regions.}
    \label{fig:en_8pi}
  \end{figure}

  Figure~\ref{fig:en_8pi} shows the average energy levels at $t=8\pi,$ i.e. for 4 revolutions of the primaries. The similarity with Fig.~(\ref{fig:en_2pi}) is obvious but we have several new splitting curves inside the stable regions, both in panel (a) and (b). In Section 4 it will be shown that these resonances correspond to the type $A$ and $D$ secondary resonances defined in \citet{Erd2}.
  
  \section{Resonant frequencies}
  
  It was shown by \citet{Erd2} that the fine structure of the $\mu-e$ plane is organized by secondary resonances between the four frequencies in the ERTBP. Two of them, the short $n_s$ and long $n_l$ frequencies (corresponding to the short and long periodic librations around $L_4$ and $L_5$), are present also in the circular problem, the additional two (in normalized units: $1-n_s,1-n_l$) appear when the primaries revolve on elliptic orbits. 
  
  In general case, Hill's equations \eqref{hill} can be considered as of two independent harmonic oscillators with periodic coefficients, i.e. with variable frequencies. Eqs~\eqref{hill} describe the two fundamental frequencies in ERTBP, namely the frequencies corresponding to the short and long periodic components of the libration around $L_4$ and $L_5.$ When the problem is circular, Eqs.~\eqref{hill} become simple harmonic oscillators and serve the well-known frequency diagram of the problem. (See Appendix A, Fig.~\ref{appfig:freqs}.)
  
  The ERM applied to the ERTBP can be used to identify the secondary resonances between the librational frequencies of the test particle and the primaries' motion. These are the resonant curves on the stability chart defined as A-type $(1-n_l):n_l$ and D-type $n_s:(1-n_s)$ in \citet{Erd2}. Investigating the problem through Eqs.~\eqref{hill} one do not need to involve the two ''additional'' normalized frequencies coming from ellipticity. Let us consider a quick example. The A11 resonance in \citet{Erd2} corresponds to the nearly V-shape boundary line of the stability zone. The notation A11 comes from $(1-n_l):n_l=1:1.$ However, after a simple rearrangement we get $1/2=n_l,$ i.e. the frequency is equal to the half of the normalized frequency of the primaries. In other words, the long period of the librating infinitesimal particle along the A11 resonance curve is twice as the primaries' period \citep{Dan}. Moreover, this rule holds for any A and D-type resonance. If we integrate Eqs.~\eqref{hill} for $t=2\pi$ and plot the sets of parameters corresponding to $E_{\rm{av}}=0$ condition, the stability map with secondary resonances, of types A and D, can be drawn from. 
  
\begin{figure}
    \centering
    \includegraphics[angle=0,width=8cm]{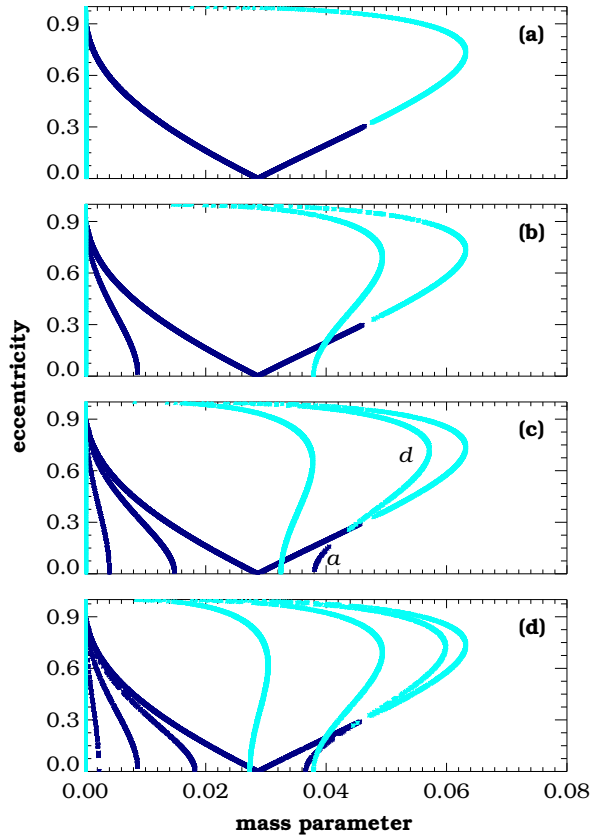}\\
    \caption{Resonant frequencies at different integration times. Computations were performed from top to bottom for times: $2\pi, 4\pi, 6\pi,\mbox{ and }8\pi.$ In panel (b)-(d) splitting curves appear. For detailed explanation see the text.}
    \label{fig:chr_diffpi}
  \end{figure}

  In Figure~\ref{fig:chr_diffpi} the $\mu-e$ plane is depicted for various integration times. Panel (a) contains the ''original'' stability chart of librational motion in the ERTBP with agreement of the energy plot of Fig.~\ref{fig:en_2pi}. Dark blue curves (color online, hereafter dark curves) correspond to the long period motion ($n_l$) while the light blue (color online, hereafter light curves) represent the short period ($n_s$). We note again, these resonant lines indicate the two components of motion around the points $L_4$ and $L_5.$ Accordingly, the nearly V-shaped dark curves describe the 2:1 commensurability of the test particle's motion, in direction $\xi_1,$ with the primaries' period. The light line represents the resonant component in the other direction $\xi_2.$ Along these curves the average energy $E_{\rm{av}},$ during one period is zero. This picture itself does not say anything about the stability of the two-dimensional motion but only the resonant frequencies. We also need the characteristics of Figs.~\ref{fig:en_2pi} and \ref{fig:en_8pi} to decide whether the motion is stable or not. Fig.~\ref{fig:chr_diffpi}(a) tells us only the border of the stability region. It has already been shown in Section 3 that libration can occur under the dark curve, where both components of Eqs.~\eqref{hill} are stable. 
  
\begin{figure}
    \centering
    \includegraphics[angle=0,width=8cm]{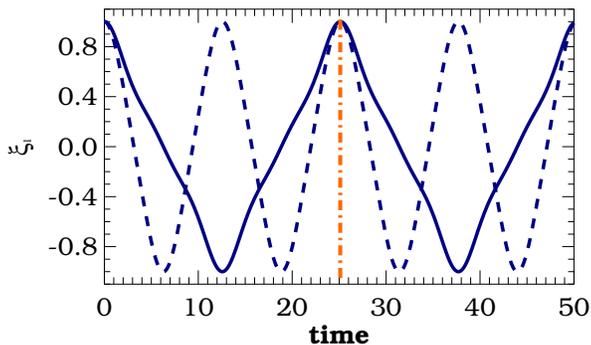}\\
    \caption{Two individual solutions of Eq.~\eqref{hilla}. The dashed line corresponds to the 2:1 resonant frequency for the parameter pair $(\mu,e)=(0.02312,0.1),$ see the circle on the dark curve in Fig.~\ref{fig:chr_diffpi}(b). While the slolid curve denotes the 4:1 resonance for $(\mu,e)=(0.00838,0.1)$, this latter point is marked with a cross in the stability chart, see also Fig.~\ref{fig:chr_diffpi}(b). The dashed and dotted line represents t=25.12 (8$\pi$), i.e. 4 revolutions of the primaries.}
    \label{fig:4to1mmr}
  \end{figure}

  Other secondary resonances show up on the stability map when the integration time is doubled, as in Fig.~\ref{fig:chr_diffpi}(b). In this panel two extra lines appear, beside the well-known stability border, corresponding to the 4:1 mean motion resonances . The plot was obtained as follows: the integration time was $4\pi,$ two periods of the primaries, the $(\mu,e)$ pairs were stored when the energy rate was zero during this time. Now, the \textit{harmonic resonant curves} are the new lines and the old borders are the \textit{sub-harmonics} of them (2:1 res.). This behaviour is defined as \textit{splitting} in the ERM, where resonant curves split the stability regions into two parts \citep{Jaz}. The 4:1 resonant curve originating from $(\mu,e)=(0.0087,0.0)$ corresponds to A31 resonance in \citet{Erd2}. To demonstrate the feature of the resonances Fig.~\ref{fig:4to1mmr} shows two individual solutions of A-type resonances, 2:1 and 4:1.

  If one continues the procedure, more and more splitting curves turn up on the stability map. In panel (c) ($t_{int}=6\pi$, 3 revolutions of the primaries) the fundamental resonances are 6:1 -- $(\mu,e)=(0.00397,0.0)$ (A51 in \citet{Erd2}) and $(\mu,e)=(0.0369,0.0)$ for type D resonance. The sub-harmonics describe the 3:1 (A21) and 2:1 (A11) resonances. 
   
  In addition two interesting curves appear in panel (c). Curve \textit{a} originates from $(\mu,e)=(0.0381,0.0),$ curve \textit{d} comes from the stability region of $\xi_2$ and terminates where curve \textit{a} ends, about at $(\mu,e)=(0.042,0.202).$ Integrating Eqs.~\eqref{hill} along these curves, one can find that these librations correspond to the \textit{super-harmonics} of the resonances (6:1). These, in question, are the 2:3 commensurabilities. The role of the meeting point is discussed in Appendix B.
  
  Figure~\ref{fig:chr_diffpi}(d) demonstrates the resonant frequencies for $t_{int}=4\mbox{ revolutions}$ of the primaries $(8\pi).$ As one may expect various new resonances appeared on the $\mu-e$ plane. Table 1 collects the resonances corresponding to dark curves on panel (d). D-type resonances are light curves of similar types.
  \begin{table}
    \caption{A-type resonances in Figure~\ref{fig:chr_diffpi}(d).}
    \label{symbols}
    \begin{tabular}{@{}lll}
      \hline
      Resonance & starting point $\mu$& panel(s)\\\hline
      8:1&0.00228&\\
      4:1&0.00876&b\\
      8:3&0.01823&\\
      2:1&0.02859&a,b,c\\
      8:5&0.03660&\\
      \hline
    \end{tabular}
    
    \medskip Second column contains the mass parameters along the $e=0$ axis where the resonant lines originate. Third column gives the other panels where the same resonance appears.
  \end{table}
 
\begin{figure}
    \centering
    \includegraphics[angle=0,width=8cm]{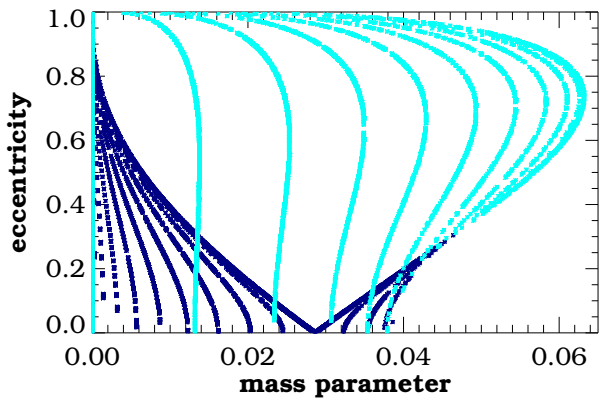}\\
    \caption{Stability chart for $t=20\pi.$ This plot contains the 10:1 resonant curves and all sub- and super-harmonics of them.}
    \label{fig:chr_20pi}
  \end{figure}

  To show how complex the stability chart can be, the $\mu-e$ plane for $t_{int}=20\pi$ is plotted in Figure~\ref{fig:chr_20pi}. The 10:1 resonances and all their sub- and super-harmonics are depicted. An even more complete and dense structure could be obtained by overplotting the charts corresponding to different integration times.
  
  \section{Summary and conclusions}
  
  In the present work the stability analysis of the ERTBP was performed. Using Hill's equations we showed that the problem can be described as two independent harmonic oscillators with time dependent frequencies. It was also demonstrated that using this approach, combined with the ERM, one can find the same stability chart as many authors published previously. 
  
  Beside many advantages, the energy-rate method allows us to obtain the resonant curves on the $\mu-e$ plane with arbitrary accuracy $\epsilon$ (obviously depending on numerical truncation errors). Moreover, it is not restricted to moderate values of the parameters as common perturbation technics.
  
  Investigating the librational motion around the $L_4$ and $L_5$ points in the ERTBP by two independent Hill equations, two unresolved problems in stability analysis can be interpreted
  \begin{enumerate}
  \item Figs.~\ref{fig:en_2pi} and \ref{fig:en_8pi} indicate that the long period component (frequencies from Eq.~\eqref{hilla}) of libration is responsible for the global feature of stability. Indeed, the A-type resonances are situated in the intersection part of the two stable regions (long and short period librations), they are therefore the backbone of the two-dimensional stable motion. 
  \item \citet{Raj} showed that there are significant peaks in the unstable domain responsible for longer escape times. In other words, there may exist some processes that stretches the lifetime of the test particle around $L_{4}$ and $L_{5}.$ Comparing their result (\citet{Raj}, Fig.~1) with Figure~\ref{fig:chr_diffpi}, one can see that the longer escape times appear exactly where the resonant frequency curves, corresponding to the shorter frequencies, are situated. An explanation of this behaviour, based on the present work, can be that the resonant curves in the stable domain of $\xi_2,$ that are actually invisible on the classical $\mu-e$ stability chart, may partially stabilize the whole librational motion around the triangular Lagrangian points.
  \end{enumerate}
  
  A direct consequence of this study can be, in future, to extend the investigation to the spatial problem, considering the third equation of motion. Interestingly, the third variational equation beside the $x$ and $y$ components of \eqref{ertbp} is already in form of Hill's equation, therefore, only a simple variable change is needed to study the extended ERTBP. 
  
  The reader may ask why are the B-type resonances \citep{Erd2} omitted from the present study. Since the ERM cannot be used to identify the exact solution of Hill's equations \eqref{hill} and therefore one cannot compare the individual frequencies, $n_s:n_l$ (perhaps the most important) resonances were not examined in this work thoroughly. However, an interesting analytical approach of B11 resonance is presented in Appendix B. 
  \section*{Acknowledgments}
  
  The author wishes to thank Prof. B. \'Erdi for helpful discussions. This work was supported by the Hungarian OTKA Grant project number K-81373.
  
  
  \appendix
  
  \section[]{Hill equations in the RTBP}
  Following Meire's \citep{Mei2} derivation, the final form of Hill's equations in the restricted three-body problem are
  \begin{equation}
    \label{apphill}
    \xi_{i}''+J_{i}(v;e,\mu)\xi_{i}=0,\hspace{1cm}i=(1,2)\\
  \end{equation}
  where
  \begin{equation*}
    \label{apphillaux1}
    \begin{split}
      J_{i}&=-\left\{rc_{1}+2-\frac{3Q_{i}+c_{2}}{q_{12}^{(i)}}+3\left[\frac{q_{22}^{(i)}}{q_{12}^{(i)}}\right]^{2}\right\},\\
      c_{i}&=1.5\left[1+(-1)^{i}\sqrt{1-g}\right]\\
      g&=3\mu(1-\mu),\\
      \mu&=\frac{m_{2}}{m_{1}+m_{2}},\\
      Q_{i}&=0.5\left[1+(-1)^{i}c+3e\cos v\right],\\
      c&=\left(1-9g+2e^2+k^2e^4\right)^{1/2},\\
      k^2&=(1-g)^{-1},\\
      q_{12}^{(1)}&=q_{12}^{(2)}-\frac{rc}{2}=a_2+e\cos v-\frac{1}{4}ke^2\cos 2v,\\
      q_{22}^{(1)}&=q_{22}^{(2)}=-\frac{1}{2}e\sin v(1-ke\cos v),\\
    \end{split}
  \end{equation*}
  with
  \begin{equation*}
    \label{apphillaux2}
    \begin{split}
      a_2=\frac{1}{4}(2c_2+1-c).
    \end{split}
  \end{equation*}
  In Eqs.~\eqref{apphill} $e$ denotes the eccentricity, $v$ is the true anomaly, $m_1$ and $m_2$ are the masses of the primaries, and $r=(1+e\cos v)^{-1}$ is the mutual distance between them.
  
  One can obtain the motion around $L_4$ and $L_5$ as a composition of two harmonic oscillations governed by Eqs.~\eqref{apphill}. Obviously, the form of equations shows that the frequencies of the motions are $\sqrt{J_i}.$ However, the solution is trivial only in the case when the frequencies do not depend on time, i.e. the orbital parameter is zero. After some computations, one can easily verify that in this case ($e=0$) the frequencies $\omega_i^2=J_i$ are
  
  \begin{equation}    
    \label{appfreqs}
    \omega_i^2=A(\mu)-B_{i}(\mu),
  \end{equation}
  where
  \begin{equation*}    
    \label{appfreqs_aux}
    \begin{split}
      A(\mu)&=2+\frac{3}{2}\left[1-\sqrt{1-g}\right],\\
      B_i(\mu)&=\frac{3+\frac{3}{2}\left(\sqrt{1-g}+(-1)^i\sqrt{1-9g}\right)}{1+\frac{3}{4}\sqrt{1-g}+(-1)^i\frac{1}{4}\sqrt{1-9g}}.
    \end{split}
  \end{equation*}
  
  \begin{figure}
    \centering
    \includegraphics[angle=0,width=8cm]{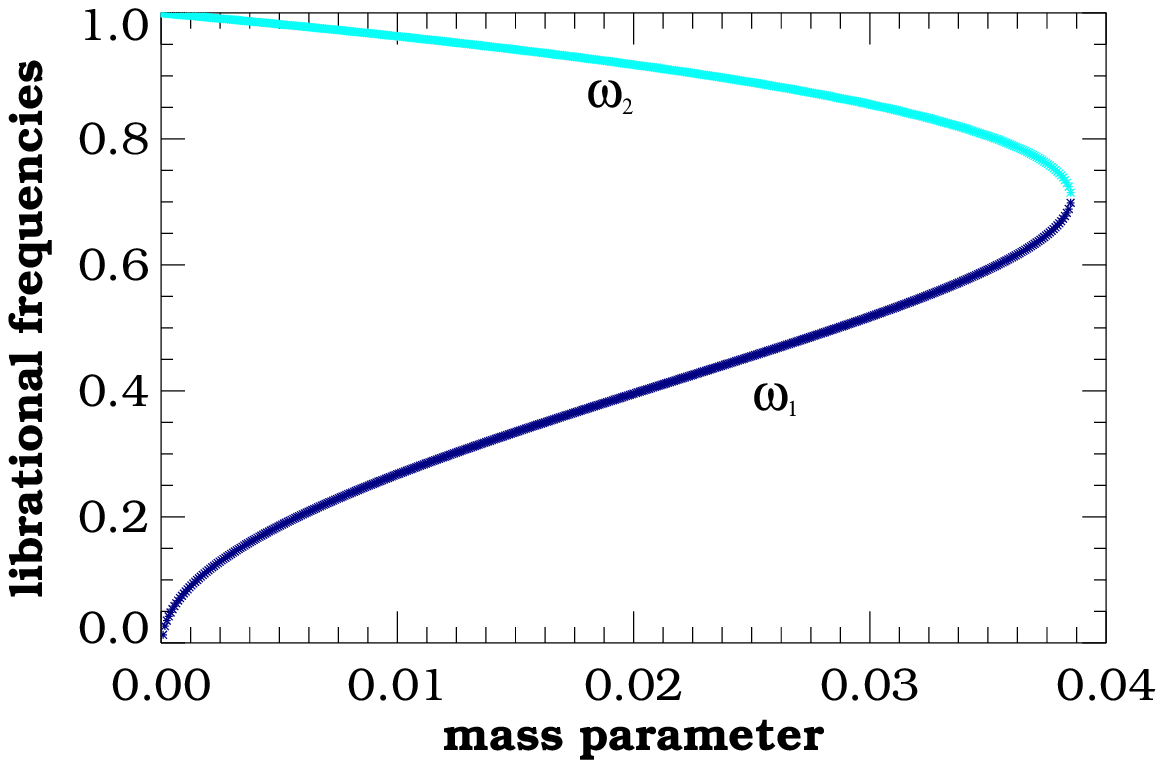}
    \caption{Librational frequencies in the circular RTBP. The dark blue (online) curve corresponds to the long period solution $\omega_1$ or $n_l.$ The light blue (online) curve depicts the short period component $\omega_2$ or $n_s.$ Eqs.~\eqref{appfreqs} provide the exact frequencies depending only on the mass parameter as $e=0$.}
    \label{appfig:freqs}
  \end{figure}
  
  Figure~\ref{appfig:freqs} demonstrates $\omega_1$ and $\omega_2$ vs. the mass parameter $\mu.$ It can be seen that the curves exactly match the well-known classical results of the circular RTBP. The lower frequency $\omega_1$ represents the long period solution while $\omega_2$ correspondes to short time oscillations. The latter tends to the frequency of the primaries (here normalized to 1) in the limit case when $\mu\to 0.$
  
  
  \section{The B11 resonance}
  In the elliptic case the frequencies of libration Eqs.~\eqref{apphill} are not so obvious. In this case we have second order linear differential equations with time dependent coefficients. For an arbitrary given periodic coefficient, there are no general procedures to express the characteristic exponents, unless the fundamental matrix of the problem is obtained by quadratures \citep{Arn,Yos}. Therefore, it seems to be a hard task to find any analytical expression for the period, and therefore the frequency $\omega_{i},$ of the solution $\xi_i(v).$ 
  
  \begin{figure}
    \centering
    \includegraphics[angle=0,width=8cm]{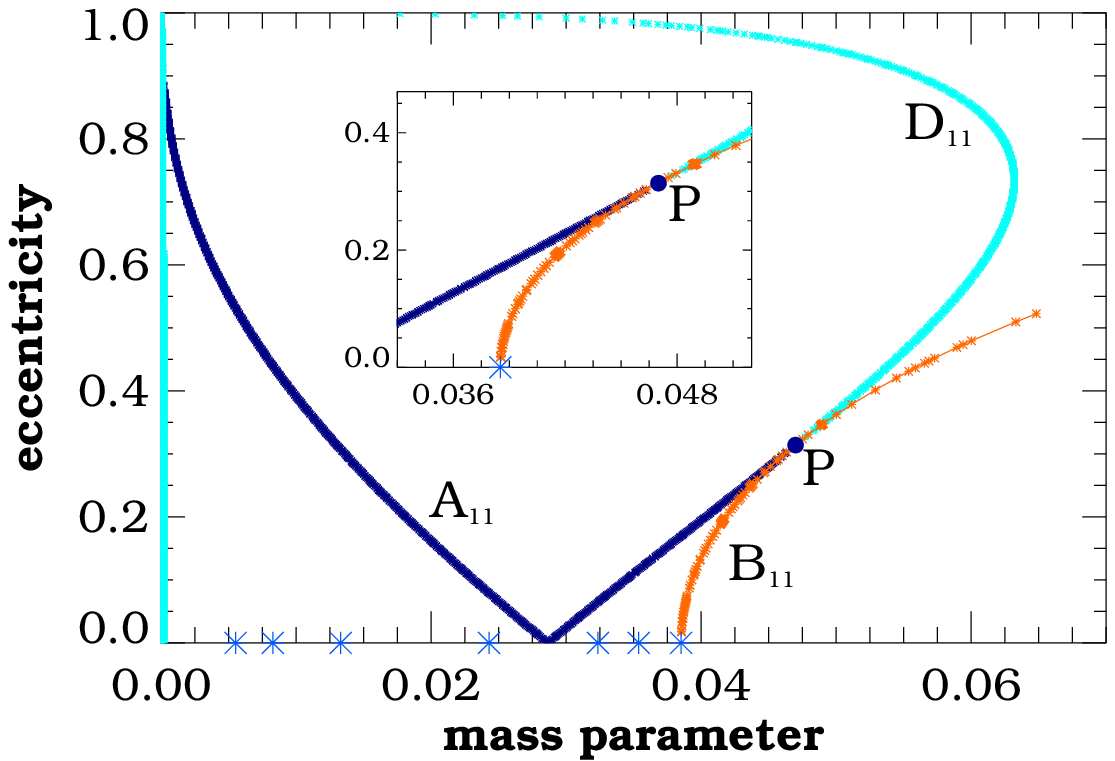}
    \caption{The $n_s:n_l=1:1$ frequency curve (B11) as the border of stability. A simple averaging by using Eqs.~\eqref{average} gives the exact curve of the B11 resonance. This curve indicates the parameter pairs $(\mu,e)$ where $\overline{\omega_1}=\overline{\omega_2}.$ The resonant curves A11, D11, and B11 meet at one point P. Additionally, several resonant points can be identified for different mass parameters at $e=0$. Crosses along the $\mu$-axis denote the following B type resonances, from left to right: 5:1, 4:1, 3:1, 2:1, 3:2, 4:3, 1:1, respectively.}
    \label{appfig:b11_res}
  \end{figure}
  Let us consider, however, a \textit{naive analytical approximation} to find the resonant curves corresponding to the B-type resonances. The B-type resonances are defined as: $n_s:n_l,$ where $n_s$ and $n_l$ denote the short and long period frequencies, respectively \citep{Erd2}. These frequencies, that are actually varying with time, can be easily obtained by evaluating the $J_i$ coefficients in Eqs.~\eqref{apphill}. \textit{Suppose} one can describe the solution with frequency $\omega_{i}$ changing around the average frequency $\overline{\omega_i}$ 
  \begin{equation}
    \label{average}
    \overline{\omega_i}=\frac{1}{T}\int\limits_{0}^{T}\sqrt{J_i(v;\mu,e)}\,\mathrm{d}v\approx\frac{1}{N}\sum\limits_{n}^{N}\sqrt{J_i(hn;\mu,e)}
  \end{equation}
  where $T$ is the period of $J_i,$ $h$ is a stepsize, and $N$ is the number of terms during one period. Consider these averaged frequencies the skeleton of individual librations around $L_4$ and $L_5.$ In this sense, B-type resonances can be evaluated simply analytically by using the formulae given by Eqs.~\eqref{average}.
  
  Just for the interesting coincidence with \'Erdi's work, who used Fourier analysis to find the exact period, the B11, i.e. $n_s:n_l=1:1,$ resonance is plotted in Figure~\ref{appfig:b11_res}. As is well-seen the curve originates at $(\mu,e)=(0.0385,0.0)$ and follows perfectly the border of stability, touching the critical point $(\mu,e)=(0.0414,0.314)$ and draw out the border line, seen also in Fig.~\ref{fig:en_2pi}, that separates two different kinds of solutions (Eqs.~\eqref{apphill}). No analytical solution in linear theory was found so far that can describe this behaviour \citep{Szeb}. 
  
  \begin{figure}
    \centering
    \includegraphics[angle=0,width=8cm]{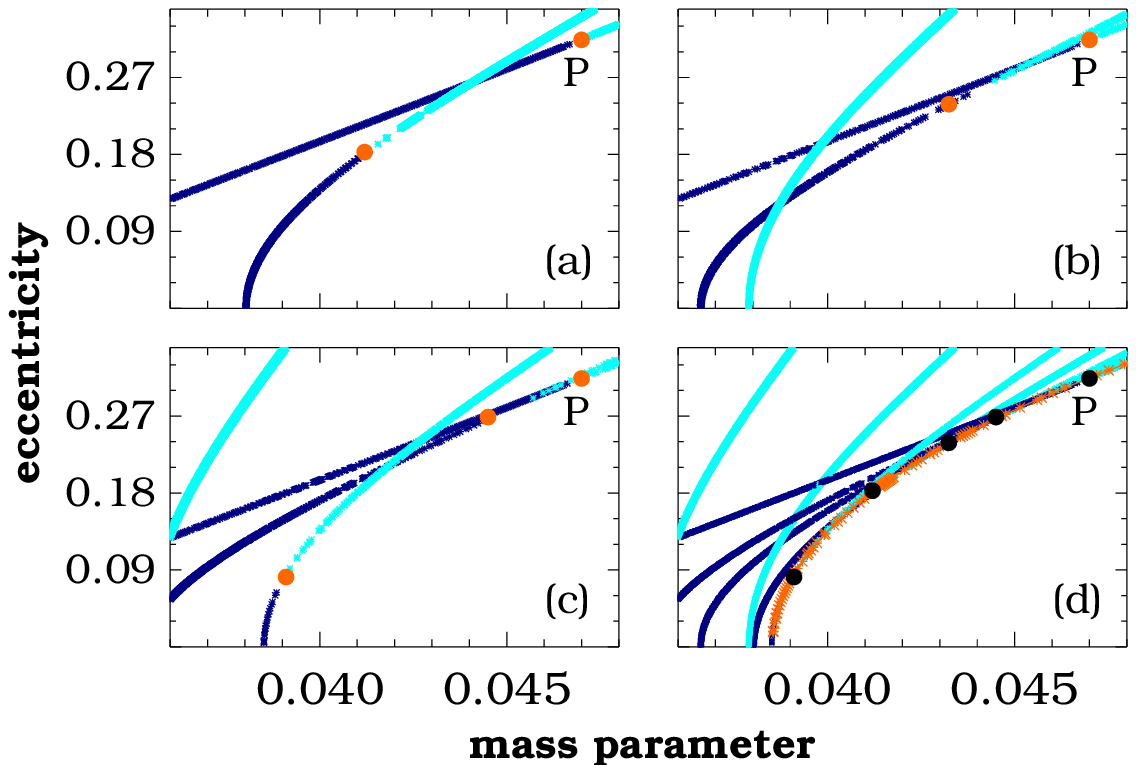}
    \caption{Various A and D-type resonances near to B11 resonance. The panels (a)-(c) refer to different integration times, $6\pi,$ $8\pi,$ $10\pi.$ (Magnified parts of Figs.~\ref{fig:chr_diffpi}(c), \ref{fig:chr_diffpi}(d), and \ref{fig:chr_20pi}, respectively.) On each panel the touching point is marked with a circle. At these points the two librational frequencies $n_s$ and $n_l$ are equal. Panel (d) verifies that the B11 resonance obtained by using Eqs.~\eqref{average} contains the circles.}
    \label{appfig:b11_magn}
  \end{figure}
  
  At this point we should explain the role of the pair of the A11 and D11 (P in Fig.~\ref{appfig:b11_res}) resonant curves that touch each other in a common point. These A and D-type resonances describe a well-defined secondary resonance between the librating particle and the primaries' motion. However, at that particular point $P(\mu,e),$ where they meet, the two individual libration frequencies become equal. Therefore, the motion at this point also satisfies the condition of B11 resonance, i.e. $\omega_{1}=\omega_{2}.$ For different integration times new A and D-type resonant curves appear on the $\mu-e$ plane that meet in one common point, see Fig.~\ref{appfig:b11_magn}. These common points are necessarily situated along the B11 resonance curve due to the condition $\omega_{1}=\omega_{2}.$
  
  Finally, we also note that this method provides the right places of the secondary resonances for the circular RTBP. That is, all the resonant frequencies (with A,B,C,D,E,F-types, \citep{Erd2}) can be identified by the analytical computation Eqs.~\eqref{appfreqs} along the $\mu$-axis, see Fig.~\ref{appfig:b11_res}. The reason of this is that the coefficients in Eqs.~\eqref{apphill} no longer depend on time in the case $e=0,$ consequently $J_i$ describe the exactly constant frequencies of the librational components, since in the circular case $J_i=\overline{\omega_i^2}.$
  
  \bsp
  
  \label{lastpage}
\end{document}